\documentclass[a4paper]{jpconf}
\usepackage{graphicx}
\usepackage{amsmath}
\usepackage{iopams}
\usepackage{bm}
\usepackage{hyperref}
\hypersetup{colorlinks,linkcolor=blue,filecolor=green,urlcolor=blue,citecolor=blue}

\newcommand{\wo}{\omega_0}
\newcommand{\wk}{{\omega_k}}
\newcommand{\wj}{{\omega_j}}
\newcommand{\wM}{{\omega_M}}
\newcommand{\wl}{{\omega_\ell}}
\newcommand{\wn}{{\omega_n}}
\newcommand{\ak}{{a_k}}
\newcommand{\akd}{{a_k^\dagger}}
\newcommand{\aj}{{a_j}}
\newcommand{\ajd}{{a_j^\dagger}}
\newcommand{\ck}{{c_k}}
\newcommand{\ckd}{{c_k^\dagger}}
\newcommand{\cj}{{c_j}}
\newcommand{\cjd}{{c_j^\dagger}}

\begin{document}
\title{Field observables near a fluctuating boundary}

\author{Federico Armata$^1$, Salvatore Butera$^2$, Federico Montalbano$^1$, Roberto Passante$^{1,3}$ and Lucia Rizzuto$^{1,3}$}
\address{$^1$ Dipartimento di Fisica e Chimica - Emilio Segr\`{e}, Universit\`{a} degli Studi di Palermo, Via Archirafi 36, I-90123 Palermo, Italia}
\address{$^2$ School of Physics and Astronomy, University of Glasgow, Glasgow G12 8QQ, United Kingdom}
\address{$^3$ INFN, Laboratori Nazionali del Sud, I-95123 Catania, Italia}

\ead{roberto.passante@unipa.it}

\begin{abstract}
We review several aspects related to the confinement of a massless scalar field in a cavity with a movable conducting wall of finite mass, free to move around its equilibrium position to which it is bound  by a harmonic potential, and whose mechanical degrees of freedom are described quantum mechanically. This system, for small displacements of the movable wall from its equilibrium position, can be described by an effective interaction Hamiltonian between the field and the mirror, quadratic in the field operators and linear in the mirror operators. In the interacting, i.e. dressed, ground state, we first consider local field observables such as the field energy density: we evaluate changes of the field energy density in the cavity with the movable wall with respect to the case of a fixed wall, and corrections to the usual Casimir forces between the two walls. We then investigate the case of two one-dimensional cavities separated by a movable wall of finite mass, with two massless scalar fields defined in the two cavities. We show that in this case correlations between the squared fields in the two cavities exist, mediated by the movable wall, at variance with the fixed-wall case.
\end{abstract}

\section{Introduction}
The presence of a cavity or reflecting boundaries has deep consequences on quantum fields, because they impose boundary conditions to the field operators. Well-known physical consequences are, for example, modifications of the spontaneous emission rate of atoms \cite{Purcell46,Meschede92,Milonni94}, the Casimir effect \cite{Casimir48,Bordag-Mohideen11} and the Casimir-Polder interactions \cite{Casimir-Polder48,Messina-Passante08}. Additional effects are present in the case of moving boundaries, for example when one boundary of a cavity is put in a nonadiabatic oscillatory motion, or its magnetodielectric properties change with time: a well-known example is the dynamical Casimir effect, consisting in the emission of pairs of real photons from the vacuum \cite{Moore70,Dodonov10,Dodonov20,Mantinan-Mazzitelli23}.

In this paper we review some recent results on observable quantities of a quantum field near a reflecting boundary with a finite mass that is free to move and whose mechanical degrees of freedom are treated quantum mechanically. Specifically, our system consists of a massless one-dimensional scalar field in a cavity with a movable boundary. The movable boundary is assumed perfectly reflecting and bound to its equilibrium position by a harmonic potential; its mechanical degrees of freedom are treated quantum mechanically, and thus it is subjected to position fluctuations. This introduces an effective field-mirror interaction, as well as an effective interaction between the field modes mediated by the movable mirror \cite{Law95,Law94}. Also, the effective field-mirror interaction yields virtual excitations of the field in the interacting ground state, i.e. a sort of dressing of the movable wall. Changes of local field quantities such as its energy density, and spatial correlations between field observables, are also expected due to the presence of the trembling cavity wall. We show that, indeed, local field quantities in the dressed vacuum state, such as the expectation value of the squared field or the field energy density, are modified due to the motion of the wall, as well as the existence of spatial correlations of the squared field between points at the opposite sides of the movable wall.

This paper is organized as follows. In Sec. \ref{sec:GroundState} we introduce our physical system of a cavity with a movable wall, its Hamiltonian and obtain the interacting ground state at the first order; we then generalize the model to the case of two cavities separated by a movable wall, and evaluate the interacting ground state at the second order. In Sec. \ref{sec:FieldObservables} we first evaluate the field energy density inside the cavity; then, in the two-cavity case, we evaluate the spatial correlation between field observables in the two cavities, in particular the correlation of the squared fields, showing that an anticorrelation exists. Possible observability of these phenomena is also discussed. Sec. \ref{sec:conclusion} is finally devoted to our final conclusions.

\section{The Hamiltonian model and the interacting ground state}
\label{sec:GroundState}
We consider a massless scalar field in a one-dimensional cavity of length $L$, with a fixed wall at $x=0$ and a movable wall at the average position $x=L$. We also assume that the movable mirror has a finite mass $m$ and that it is bound to its average position by a harmonic potential of angular frequency $\wo$. Both mirrors are assumed to be perfectly reflecting. The mechanical degrees of freedom of the movable mirror are described quantum mechanically, in terms of a quantum harmonic oscillator.

In the case of small displacements of the movable mirror from its equilibrium position, our system can be described by an effective Hamiltonian originally introduced by Law \cite{Law95}. This effective Hamiltonian $H$ contains an unperturbed term given by the sum of the field and mirror Hamiltonians and an interaction term that is quadratic in the field operators and linear in the mirror operators. The field annihilation and creation operators refer to the movable wall's equilibrium position. The Law Hamiltonian is \cite{Law95}
\begin{equation}
\label{Hamiltonian1}
H=\sum_k \hbar \wk \akd \ak +\hbar \wo b^\dagger b -(b+b^\dagger )\sum_{kj} C_{kj} N\left[ (\ak +\akd)(\aj +\ajd)\right] ,
\end{equation}
where $b$ and $b^\dagger$ are bosonic annihilation and creation operators of the movable wall, $\ak$ and $\akd$ are the bosonic operators of the massless 1D scalar field relative to the wall's equilibrium position, $\wk = ck$, $N$ is the normal ordering operator, and $C_{kj}=(-1)^{k+j} L^{-1} \sqrt{\hbar^3\wk \wj/(8m\wo)}$ is the field-mirror coupling constant. This kind of Hamiltonian has been often used to treat the mirror-field dynamics \cite{Cheung-Law11,Armata-Kim17,Butera22}. In the following, we will treat the field-mirror interaction term perturbatively, up to the first or second order according to the field quantities we shall consider.

The unperturbed ground state (both mirror and field in their bare vacuum state) is $\lvert 0; \{ 0_k\} \rangle$, where the first element is relative to the movable wall, and the second to the field. The presence of the effective mirror-field interaction yields an energy shift of the ground state given, at the second order in the coupling, by \cite{Butera-Passante13}
\begin{equation}
\label{energyshift}
\Delta E_g = - \frac {\hbar^2}{4L^2m\wo} \sum_{kj} \frac {\wk \wj}{\wo +\wk +\wj} .
\end{equation}

We now consider the system's interacting ground state. At first order in the coupling constant $C_{kj}$, the corrected ground state is given by (apart a normalization factor)
\begin{equation}
\label{groundstate1}
\lvert g \rangle = \lvert 0; \{ 0_k\} \rangle + \frac 1L \left( \frac {\hbar}{8m\wo} \right)^{1/2} \sum_{kj} (-1)^{k+j}  \frac {\sqrt{\wk \wj}}{\wo +\wk +\wj} \lvert 1; \{ 1_k 1_j \} \rangle ,
\end{equation}
that contains virtual admixtures with states with one excitation in the wall's degrees of freedom and two excitations in the field. From the state (\ref{groundstate1}), it is possible to evaluate the virtual photon spectrum, showing that it has a maximum when the sum of the frequencies of the two photons is equal to the oscillation frequency $\wo$ of the boundary; this is analogous to the case of the dynamical Casimir effect (forced harmonic oscillation of the boundary), except that in the latter case the condition $\wk +\wj =\wo$ is sharp, while in our present case it is only a maximum of a quite broad frequency distribution.

In the next section, we will also consider the case of two 1D cavities of length $L$ separated by a perfectly reflecting movable wall, and two massless quantum scalar fields defined inside each cavity. If the perfect boundary separating the two cavities were fixed in space, the fields in the two cavities cannot influence each other, of course; this means that the spatial correlation between field observables pertinent to the different cavities are vanishing. The situation changes drastically if the separating wall can move, as we shall discuss in detail in the next section.

We now introduce the Hamiltonian model for this two-cavity system that is a straightforward generalization of the previous one-cavity model. The average position of the movable wall, that separates the two cavities, is at $x=L$, while the two fixed walls are at $x=0$ (cavity 1, at the left side of the movable wall) and at $x=2L$ (cavity 2, at the right side of the movable wall). The Hamiltonian of our system, as mentioned, is a straightforward generalization of the Hamiltonian (\ref{Hamiltonian1}) to the present two-cavity case; it is given by
\begin{eqnarray}
\label{Hamiltonian2}
H &=& \hbar \sum _k \wk \akd \ak + \hbar \sum _k \wk \ckd \ck + \hbar \wo b^\dagger b
-\left( b+b^\dagger \right) \sum_{kj} C_{kj}^1 \text{N} \left[ \left( \aj + \ajd \right) \left( \ak + \akd \right) \right]
\nonumber \\
&\ &  -\left( b+b^\dagger \right) \sum_{kj} C_{kj}^2 \text{N} \left[ \left( \cj + \cjd \right) \left( \ck + \ckd \right) \right] ,
 \end{eqnarray}
where $\ak$ and $\akd$ are bosonic operators relative to the field in the left-side cavity (cavity 1), while $\ck$ and $\ckd$ are the bosonic operators relative to the right-side cavity (cavity 2). They both refer to modes relative to the equilibrium position $x=L$ of the mobile wall. There are also two mirror-field interaction terms, with $C_{kj}^1$ and $C_{kj}^2$, respectively, the coupling constants of the mirror with the field defined in cavity 1 and in cavity 2. Also, $C_{kj}^1$ is equal to the coupling constant $C_{kj}$ defined after Eq. (\ref{Hamiltonian1}) for the single-cavity case, and $C_{kj}^2 = - C_{kj}^1$, because a movement of the movable wall in one direction increases the length of one cavity and decreases the length of the other one \cite{Armata-Kim17,Montalbano-Armata23}.

The noninteracting (bare) ground state is $\lvert 0 ; \{ 0_k\} ;  \{ 0_k\}  \rangle$, where the first element refers to the mirror's excitations, while the second  and the third ones refer respectively to the quanta in cavity 1 and in cavity 2: in this state there are no excitations in the mirror and in any of the fields. The interaction terms in (\ref{Hamiltonian2}) give a correction yielding the interacting (dressed) state. Up to the second order in the coupling constants (we need the second-order corrected state because the correlation function we are going to calculate in Sec. \ref{sec:FieldObservables} is nonvanishing starting from the second order in the couplings), the true ground state has the following form
\begin{equation}
\label{corrected state}
\lvert \tilde{g} \rangle = \left( 1 -\frac 12 \Lambda^2 \right)\lvert 0;  \{ 0_k \}  ; \{ 0_k \} \rangle + \lvert g^{(1)} \rangle +  \lvert g^{(2)} \rangle ,
\end{equation}
where $\Lambda$ is a normalization factor, $ \lvert g^{(1)} \rangle$ is the first-order correction and $\lvert g^{(2)} \rangle$ is the second-order correction. Using time-independent perturbation theory up to the second order, we obtain combinations of states of the form described in the following. The first order correction $\lvert g^{(1)} \rangle$ contains states of the form $\lvert 1; \{ 1_j 1_k \} ;  \{ 0_k\} \rangle$ (one mirror excitation, two quanta in cavity 1 and zero in cavity 2) or $\lvert 1;  \{ 0_k\};  \{ 1_j 1_k \} \rangle$ (one mirror excitation, zero quanta in cavity 1 and two quanta in cavity 2). The second-order correction $\lvert g^{(2)} \rangle$ contains terms with zero mirror excitations and one of the following possibilities for the two fields:  two quanta in one cavity and zero quanta in the other one, or two quanta in each cavity, or four quanta in one cavity and zero quanta in the other cavity (there are also states with two mirror excitations, but they do not contribute to the correlation functions we are going to calculate in Sec. \ref{sec:FieldObservables}) \cite{Montalbano-Armata23}.

\section{Field observables near the movable wall for the one- and two-cavity system}
\label{sec:FieldObservables}

We first consider the one-cavity case. A relevant aspect is considering local field quantities inside the cavity such as the field energy density ${\cal H}(x)= \frac 12 \left[ \frac 1{c^2} \phi^2(x)+\left( \frac {d\phi (x)}{dx}\right)^2\right]$. The expression of the 1D scalar field operator, with the appropriate boundary conditions at $x=0$ and $x=L$, is
\begin{equation}
\label{fieldoperator}
\phi (x) = \sqrt{\frac {\hbar c^2}{L}} \sum_j \frac {\sin (k_jx)}{\sqrt{\wj}} \left( \aj +\ajd \right) ,
\end{equation}
where $k_j = n_j \pi /L$, with $n_j=1,2,...$, and $\wj = ck_j$. We now evaluate the expression for the change of the renormalized field energy density inside the cavity, evaluated on the interacting ground state (\ref{groundstate1}), with respect to that for a cavity with fixed walls $\Delta {\cal H}(x) = \langle g \rvert {\cal H}(x) \lvert g \rangle - \langle 0 \rvert {\cal H}(x) \lvert 0 \rangle$. The result is
\begin{equation}
\label{energydensity}
\Delta {\cal H}(x) = \frac{\hbar^2}{2L^3m\wo} \sum_{jk\ell} (-1)^{k+\ell} \frac {\wj \wk \wl}{(\wo +\wj +\wk )(\wo +\wj +\wl)} \cos \left[ \frac {(\wk-\omega_\ell )x}{c} \right] f_{kj\ell}(\wM ),
\end{equation}
where $x$ is the distance from the movable wall and $f_{kj\ell}(\wM )$ is a regularization function necessary to cure ultraviolet divergences and to simulate the effect of a real metal, $\wM$ being an upper cutoff frequency. In Ref. \cite{Butera-Passante13} $\Delta {\cal H}(x)$ was evaluated numerically using a sharp cutoff function; this, for a finite cavity size $L$, is equivalent to considering a finite number of field modes. The results from the numerical evaluation show that, by including the motion of the finite-mass wall, the field energy density inside the cavity changes, and that this change is particularly relevant in the proximity of the movable wall (the effect increases with increasing cutoff frequency), while it is negligible at large distances form the movable wall. This result is consistent with the following physical picture. Due to the field-mirror interaction, pairs of virtual quanta are emitted and reabsorbed by the wall (see Eq. (\ref{groundstate1})). These quanta, however, remain confined near the wall, according to the energy-time uncertainty relation: the higher their frequency is, the more are they confined near the mobile wall. It is also worth to stress that the field energy density we are considering can be probed through the atom-wall dispersion interaction energy with a polarizable body such as a ground-state atom \cite{Passante18}:  thus, the change in the field energy density inside the cavity due to the position fluctuations of the movable wall can be in principle measured. Eq. (\ref{energydensity}) also shows that the effect found becomes larger as the mass and oscillation frequency of the movable wall are decreased.

We also wish to mention that expressions like Eq. (\ref{energydensity}) could be also evaluated with an exponential cutoff function (more realistic than a sharp one), in particular in continuum limit, $L \rightarrow \infty$, $\sum_k \rightarrow L/(2\pi ) \int dk$, thus recovering the case of a single movable wall, as we will explicitly do in the next part of this section for the spatial correlations in the two-cavity case.

These results can be extended to the case of a 1D electromagnetic field or to a 3D scalar field \cite{Armata-Passante15}.  For example, in the case of the 1D electromagnetic case, the correction to the electric and magnetic field fluctuations (or, equivalently, to the electric and magnetic energy densities) in the cavity due to the wall's position fluctuations at the first order are found to be \cite{Armata-Passante15}
\begin{equation}
\label{electricenergydensity}
\langle E_z^2(x) \rangle = \frac {\hbar^2}{m\wo L^3} \sum_{j\ell n} (-1)^{\ell +n}
\frac {\wj \wl \wn  f_{j\ell n}(\wM )}{(\wo +\wj +\wl)(\wo +\wj +\wn)} \sin (k_\ell x) \sin (k_nx) ,
\end{equation}
\begin{equation}
\label{magneticenergydensity}
\langle B_y^2(x) \rangle = \frac {\hbar^2}{m\wo L^3} \sum_{j\ell n} (-1)^{\ell +n}
\frac {\wj \wl \wn f_{j\ell n}(\wM )}{(\wo +\wj +\wl)(\wo +\wj +\wn)} \cos (k_\ell x) \cos (k_nx)  ,
\end{equation}
where the subscripts in the electric and magnetic field operators refer to their cartesian components and $f_{j\ell n}(\wM )$ is an appropriate cutoff function for $\wj ,\wl ,\wn$.
The main qualitative features of the results obtained for the field energy densities of the 1D electric and magnetic fields, as well as of the 3D scalar field, are similar to the case considered before for the 1D scalar field. We refer the reader to Ref. \cite{Armata-Passante15} for more details. For example, explicit evaluation of (\ref{electricenergydensity},\ref{magneticenergydensity}) shows significant changes of the electric and magnetic energy density in the cavity, this change rapidly increasing when approaching the movable wall.

We now consider the case of two cavities separated by a perfectly reflecting movable mirror, introduced at the end of Sec. \ref{sec:GroundState}. Specifically, we calculate the spatial correlation functions between field observables defined in the two cavities on the interacting ground state (\ref{corrected state}). The field operator in cavity 1 ($x_1 \in (0,L)$) is
\begin{equation}
\label{fieldop1}
\phi (x_1) = \sqrt{\frac {\hbar c^2}{L}} \sum_j \frac {\sin (k_j x_1)}{\sqrt{\wj}} \left( \aj + \ajd \right) ,
\end{equation}
and the field operator in cavity 2 ($x_2 \in (L,2L)$) is
\begin{equation}
\label{fieldop2}
\phi (x_2) = -\sqrt{\frac {\hbar c^2}{L}} \sum_j \frac {\sin (k_j x_2)}{\sqrt{\wj}} \left( \cj + \cjd \right) ,
\end{equation}
with $k_j=n_j \pi /L$, $n_j =1,2,...$ due to the boundary conditions of the two field operators and $\wj =ck_j$.
It is immediate to see that the spatial correlation function on the dressed ground state (\ref{corrected state}) between these two fields (i.e. between a point in cavity 1 and a point in cavity 2) vanishes,
$\langle \tilde{g} \lvert \phi (x_1) \phi (x_2) \rvert \tilde{g} \rangle - \langle \tilde{g} \lvert \phi (x_1) \rvert \tilde{g} \rangle \langle \tilde{g} \lvert \phi (x_2) \rvert \tilde{g} \rangle = 0$ (this is indeed true at any order in perturbation theory, within the Hamiltonian model (\ref{Hamiltonian2}) used).

The spatial correlation of the {\it squared} fields between the two cavities, with $x_1 \in (0,L)$ and $x_2 \in (L,2L)$, evaluated on the state (\ref{corrected state}), after some lengthy algebraic calculation is obtained as \cite{Montalbano-Armata23}
\begin{eqnarray}
\label{correlationsquaredfield}
C(x_1,x_2) &=& \langle \tilde{g} \lvert \phi^2(x_1) \phi^2(x_2) \rvert \tilde{g} \rangle - \langle \tilde{g} \lvert \phi^2(x_1) \rvert \tilde{g} \rangle \langle \tilde{g} \lvert \phi^2(x_2) \rvert \tilde{g} \rangle
\nonumber \\
&=& -\frac {\hbar^3c^4}{L^4m\wo} \sum_{pqrs} (-1)^{p+q+r+s}
\Big\{ \frac {\sin (k_p x_1)\sin (k_q x_1) \sin (k_r x_2) \sin (k_s x_2)}{(\wo +\omega_p +\omega_q)(\wo +\omega_r +\omega_s)}
\nonumber \\
&\ & \ + \Big[ \frac {\sin (k_p x_1)\sin (k_q x_1) \sin (k_r x_2) \sin (k_s x_2)}{(\wo +\omega_p +\omega_q)(\omega_p +\omega_q +\omega_r +\omega_s)} + (x_1 \leftrightarrow x_2) \Big] \Big\} f_{pqrs}(\wM ) ,
\end{eqnarray}
where
\begin{equation}
\label{exponential cutoff}
f_{pqrs}(\wM ) = \text{exp}\left[ -(\omega_p +\omega_q +\omega_r +\omega_s)/\wM \right]
\end{equation}
is a regularization function and $\wM$ is the cutoff frequency; in the case of boundaries made of a real metal, the cutoff frequency $\wM$ can be also identified with the metal plasma frequency.

In the continuum limit, $L \rightarrow \infty$, $\sum_j \rightarrow (L/2\pi ) \int_0^\infty dk_j$, and defining in both cavities the distance from the movable wall as ${\tilde{x}}_1 = L-x_1$ and ${\tilde{x}}_2 = x_2 - L$, we recover the case of a single movable boundary, to which we are mainly interested. In the continuum limit and for generic values of the distances ${\tilde{x}}_1 = L-x_1$ and ${\tilde{x}}_2 = x_2 - L$, the spatial correlation function  (\ref{correlationsquaredfield}) contains four frequency integrals, three of them can be obtained analytically, and we must resort to a numerical integration for the last one. For large distances from the wall, i.e. for ${\tilde{x}}_{1,2} \gg c/\wo$, and using $\wM \gg \wo$ (condition certainly met in any realistic setup), we can however obtain an approximated analytical expression for the spatial correlation function
\begin{equation}
\label{apprcorrfunction}
C({\tilde{x}}_1,{\tilde{x}}_2) \simeq - \frac{\hbar^3c^4}{2^9 \pi^4} \frac 1{m\wo^3} \frac 1{{\tilde{x}}_1^2 {\tilde{x}}_2^2} .
\end{equation}

Equation (\ref{apprcorrfunction}) proves the existence of a nonvanishing correlation between the squared fields in the two cavities, even if the cavities are separated by a perfectly reflecting mirror.
$C({\tilde{x}}_1,{\tilde{x}}_2)$ is negative, and thus the squared fields at the opposite sides of the movable wall are anticorrelated, even if there is not any direct interaction between them. The physical origin of this result is indeed in the mutual interaction of both fields with the movable mirror. For points at the same distance $\tilde{x}= {\tilde{x}}_1 = {\tilde{x}}_2$ from the wall, the anticorrelation scales with the distance as ${\tilde{x}}^{-4}$, and, respectively, as $m^{-1}$ and $\wo^{-3}$ from the mirror's mass and oscillation frequency. Even if Eq. (\ref{apprcorrfunction}) holds only at large distances from the mirror, the numerical evaluation of (\ref{correlationsquaredfield}) shows that a nonvanishing correlation exists also at shorter distances from the movable mirror  \cite{Montalbano-Armata23}. Since, according to (\ref{correlationsquaredfield}), the (anti)correlation between the squared field scales as $1/m$ and $1/\wo^3$, the effects we have described are larger the smaller the mass and oscillation frequency of the movable wall are. In optomechanical experiments, a typical oscillation frequency is of the order of $10^4-10^6 \, {\text{s}}^{-1}$ and masses as low as $10^{-15}$ kg to $10^{-21}$ kg can be experimentally achieved \cite{Meystre13,Aspelmeyer-Kippenberg14}. These very low values of the mass could make experimentally detectable the effects we have found for the field energy density and squared field spatial correlations, for example exploiting the connection between field energy densities and two- and many-body dispersion interactions \cite{Montalbano-Armata23,Passante18}. Finally, we wish to mention that the results here presented have some similarity with the emission of radiation by a fuzzy black-hole event horizon \cite{Arias-Krein12,Takahashi-Soda10}, and we guess it could be worth to pursue more deeply this analogy.

\section{Conclusion}
\label{sec:conclusion}

In this paper, we have reviewed some relevant aspects related to quantum fields confined in one or two perfectly reflecting one-dimensional cavities with a wall of finite mass and free to move, bound to its equilibrium position by a harmonic potential. The mechanical degrees of freedom of the movable wall are treated quantum mechanical, and this yield a mirror-field interaction and an effective interaction between the field modes, mediated by the movable boundary. Using the Law Hamiltonian, in the case of a single cavity with a movable boundary, we have investigated the effect of the motion and position fluctuations of the wall on some local field observables (energy densities and squared field) in the interacting ground state, which contains virtual field excitations; we find significant changes of the field energy density inside the cavity. In the case of two ideal cavities separated by a movable wall, we have investigated the spatial correlations between the squared field in the two cavities, showing that an anticorrelation exists, notwithstanding they are separated by a perfectly reflecting mirror. We have discussed the dependence of such spatial anticorrelation from the relevant parameters, in particular from the distance from the movable wall and the wall's mass and oscillation frequency. Possible observability of these effects has been also discussed.

\ack
R.~P. and L.~R. acknowledge financial support from the Julian Schwinger Foundation. R.~P. and L.~R. also acknowledge partial financial support from the FFR2021 grant from the University of Palermo, Italy. S.~B. acknowledges funding from the Leverhulme Trust Grant No. ECF-2019-461, and from University of Glasgow via the Lord Kelvin/Adam Smith (LKAS) Leadership Fellowship

\section*{References}
\bibliography{biblio.bib}

\providecommand{\newblock}{}
\begin{thebibliography}{10}
\expandafter\ifx\csname url\endcsname\relax
  \def\url#1{{\tt #1}}\fi
\expandafter\ifx\csname urlprefix\endcsname\relax\def\urlprefix{URL }\fi
\providecommand{\eprint}[2][]{\url{#2}}

\bibitem{Purcell46}
Purcell E 1946 {\em Phys. Rev.\/} {\bf 69}(11-12) 681
  \urlprefix\url{https://link.aps.org/doi/10.1103/PhysRev.69.674}

\bibitem{Meschede92}
Meschede D 1992 {\em Physics Reports\/} {\bf 211} 201--250 ISSN 0370-1573
  \urlprefix\url{https://www.sciencedirect.com/science/article/pii/037015739290110L}

\bibitem{Milonni94}
Milonni P 1994 {\em The Quantum Vacuum: An Introduction to Quantum
  Electrodynamics\/} (San Diego, CA: Academic Press)

\bibitem{Casimir48}
Casimir H~B~G 1948 {\em Proc. Kon. Ned. Akad. Wet.\/} {\bf B51} 793

\bibitem{Bordag-Mohideen11}
Bordag M, Mohideen U and Mostepanenko V 2001 {\em Physics Reports\/} {\bf 353}
  1--205 ISSN 0370-1573
  \urlprefix\url{https://www.sciencedirect.com/science/article/pii/S0370157301000151}

\bibitem{Casimir-Polder48}
Casimir H~B~G and Polder D 1948 {\em Phys. Rev.\/} {\bf 73}(4) 360--372
  \urlprefix\url{https://link.aps.org/doi/10.1103/PhysRev.73.360}

\bibitem{Messina-Passante08}
Messina R, Passante R, Rizzuto L, Spagnolo S and Vasile R 2008 {\em Journal of
  Physics A: Mathematical and Theoretical\/} {\bf 41} 164031
  \urlprefix\url{http://stacks.iop.org/1751-8121/41/i=16/a=164031}

\bibitem{Moore70}
Moore G~T 1970 {\em Journal of Mathematical Physics\/} {\bf 11} 2679--2691
  \urlprefix\url{https://doi.org/10.1063/1.1665432}

\bibitem{Dodonov10}
Dodonov V~V 2010 {\em Physica Scripta\/} {\bf 82} 038105
  \urlprefix\url{https://doi.org/10.1088/0031-8949/82/03/038105}

\bibitem{Dodonov20}
Dodonov V 2020 {\em Physics\/} {\bf 2} 67--104 ISSN 2624-8174
  \urlprefix\url{https://www.mdpi.com/2624-8174/2/1/7}

\bibitem{Mantinan-Mazzitelli23}
Mantiñan M, Mazzitelli F~D and Trombetta L~G 2023 {\em Entropy\/} {\bf 25}
  ISSN 1099-4300 \urlprefix\url{https://www.mdpi.com/1099-4300/25/1/151}

\bibitem{Law95}
Law C~K 1995 {\em Phys. Rev. A\/} {\bf 51}(3) 2537--2541
  \urlprefix\url{https://link.aps.org/doi/10.1103/PhysRevA.51.2537}

\bibitem{Law94}
Law C~K 1994 {\em Phys. Rev. A\/} {\bf 49}(1) 433--437
  \urlprefix\url{https://link.aps.org/doi/10.1103/PhysRevA.49.433}

\bibitem{Cheung-Law11}
Cheung H~K and Law C~K 2011 {\em Phys. Rev. A\/} {\bf 84}(2) 023812
  \urlprefix\url{https://link.aps.org/doi/10.1103/PhysRevA.84.023812}

\bibitem{Armata-Kim17}
Armata F, Kim M~S, Butera S, Rizzuto L and Passante R 2017 {\em Phys. Rev. D\/}
  {\bf 96}(4) 045007
  \urlprefix\url{https://link.aps.org/doi/10.1103/PhysRevD.96.045007}

\bibitem{Butera22}
Butera S 2022 {\em Phys. Rev. D\/} {\bf 105}(1) 016023
  \urlprefix\url{https://link.aps.org/doi/10.1103/PhysRevD.105.016023}

\bibitem{Butera-Passante13}
Butera S and Passante R 2013 {\em Phys. Rev. Lett.\/} {\bf 111}(6) 060403
  \urlprefix\url{https://link.aps.org/doi/10.1103/PhysRevLett.111.060403}

\bibitem{Montalbano-Armata23}
Montalbano F, Armata F, Rizzuto L and Passante R 2023 {\em Phys. Rev. D\/} {\bf
  107}(5) 056007
  \urlprefix\url{https://link.aps.org/doi/10.1103/PhysRevD.107.056007}

\bibitem{Passante18}
Passante R 2018 {\em Symmetry\/} {\bf 10} 735 ISSN 2073-8994
  \urlprefix\url{https://www.mdpi.com/2073-8994/10/12/735}

\bibitem{Armata-Passante15}
Armata F and Passante R 2015 {\em Phys. Rev. D\/} {\bf 91}(2) 025012
  \urlprefix\url{https://link.aps.org/doi/10.1103/PhysRevD.91.025012}

\bibitem{Meystre13}
Meystre P 2013 {\em Annalen der Physik\/} {\bf 525} 215--233
  \urlprefix\url{https://onlinelibrary.wiley.com/doi/abs/10.1002/andp.201200226}

\bibitem{Aspelmeyer-Kippenberg14}
Aspelmeyer M, Kippenberg T~J and Marquardt F 2014 {\em Rev. Mod. Phys.\/} {\bf
  86}(4) 1391--1452
  \urlprefix\url{https://link.aps.org/doi/10.1103/RevModPhys.86.1391}

\bibitem{Arias-Krein12}
Arias E, Krein G, Menezes G and Svaiter N~F 2012 {\em International Journal of
  Modern Physics A\/} {\bf 27} 1250129

\bibitem{Takahashi-Soda10}
Takahashi T and Soda J 2010 {\em Classical and Quantum Gravity\/} {\bf 27}
  175008 \urlprefix\url{https://doi.org/10.1088/0264-9381/27/17/175008}

\end{thebibliography}

\end{document}